# A Robust Approach for the Growth of Epitaxial Spinel Ferrite Films

J. *X. Ma, D. Mazumdar, G. Kim, H. Sato, N. Z. Bao*, and *A. Gupta**

Center for Materials for Information Technology, 205 Bevill Building, Box 870209,

University of Alabama, Tuscaloosa, AL 35487

Abstract:

Heteroepitaxial spinel ferrites $NiFe_2O_4$ and $CoFe_2O_4$ films have been prepared by pulsed laser deposition (PLD) at various temperatures (175 - 690 °C) under ozone/oxygen pressure of 10 mTorr. Due to enhanced kinetic energy of ablated species at low pressure and enhanced oxidation power of ozone, epitaxy has been achieved at significantly lower temperatures than previously reported. Films grown at temperature below 550 °C show a novel growth mode, which we term "vertical step-flow" growth mode. Epitaxial spinel ferrite films with atomically flat surface over large areas and enhanced magnetic moment can be routinely obtained. Interestingly, the growth mode is independent of the nature of substrates (spinel $MgAl_2O_4$, perovskite $SrTiO_3$, and rock salt MgO) and film thicknesses. The underlying growth mechanism is discussed.

To whom correspondence should be addressed: agupta@mint.ua.edu

I.   Introduction

There is considerable interest in the growth of high-quality, single crystal spinel ferrites (general formula $AB_2O_4$) films because to their numerous technological applications in areas such as microwave integrated devices,[1] magnetoelectric (ME) coupling heterostructures,[2,3,4] and potentially as an active barrier material in an emerging class of spintronic devices called spin filters.[5,6,7] This is in large part due to the unique property of a large class of spinel ferrites, including $CoFe_2O_4$ (CFO) and $NiFe_2O_4$ (NFO), of being magnetic insulators with a high magnetic ordering temperature. The net magnetization in these materials is due to the existence of two magnetic sublattices, and they are thus strictly ferrimagnetic. Spinel oxides have a complex crystal structure with a large unit cell consisting of many interstitial sites. Moreover, the transition metals can adopt various oxidation states. Thus, it is a challenging task to grow epitaxial spinel films (especially thick films) with low defect density and excellent magnetic properties. Up to date, various thin film techniques, including pulsed laser deposition (PLD), sputtering, evaporation and chemical techniques have been used to grow epitaxial ferrite films.[8,9,10,11,12] But the magnetic properties of epitaxially grown spinel ferrite films are generally far from ideal in comparison to those in the bulk, particularly for as-deposited films without post-anneal. The saturation magnetic moment ($M_S$) of CFO films is generally between 120-350 emu/cm$^3$, well below the bulk value of 450 emu/cm$^3$. Suzuki *et al.* reported a $M_S$ value of 400 emu/cm$^3$ for CFO films grown on $CoCr_2O_4$-buffered $MgAl_2O_4$ substrates.[13] For NFO films, Venzke *et al.* obtained a $M_S$ value of 190 emu/cm$^3$, also quite a bit lower than the bulk value of 300 emu/cm$^3$.[14] Recently, Luders *et al.* reported significantly enhanced $M_S$ of 1050

emu/cm$^3$ in ultrathin films of 3 nm in thickness, which they attributed to stabilization of the normal spinel near the interface.[5]

One of the major obstacles in obtaining high-quality epitaxial ferrite films with bulk magnetic properties is the lack of availability of isostructural substrates with good lattice match. So far, ferrite films have primarily been grown on oxides substrates such as perovskite SrTiO$_3$ (STO),[14] rock-salt MgO,[15] isostructural spinel MgAl$_2$O$_4$ (MAO),[13] and various buffered substrates. None of these are ideal for the growth of ferrite films. MAO, with a lattice constant of 8.08 Å, has a large mismatch with NFO and CFO of over 3% ($a$ = 8.34Å for NFO and 8.38Å for CFO). On the other hand, rock-salt MgO is better lattice matched with the ferrites, with a lattice parameter almost twice that of MgO ($a$ = 4.212Å). But spinel ferrite films grown on MgO are prone to anti-phase boundaries (APBs) from cation stacking defects originating from equivalent nucleation sites. Such films do not saturate even at magnetic fields as high as 5 Tesla and ultra-thin films have been reported to be even superparamagnetic.[16] An additional problem is the inter-diffusion associated with high growth temperature of ferrite films. Inter-diffusion is an especially seriously issue for the growth of ferrite/ferroelectric heterostructures, since a number of ferroelectric materials of interest consist of volatile elements. Also, high temperature growth on large mismatched substrates results in 3D island growth mode leading to a rough surface morphology which is linked with misfit dislocation introduction.[17] In order to circumvent these issues, a low temperature epitaxial growth method that results in films with excellent structural and magnetic properties is highly desirable.

It is well known that for high-misfit heteroepitaxial growth of semiconductor such as Ge on Si, an effective approach to suppress island formation and achieve layer-by-layer growth is by reduction of the growth temperature.[18] The degree to which one can control island formation by low-temperature growth depends on the minimum temperature at which deposition leads to epitaxial growth. In the case of complex oxides including spinel ferrites, epitaxial growth is very sensitive to phase equilibrium and the oxidation kinetics, both of which are generally not favored at low temperatures. Nevertheless, Kiyomura et al.[19] have reported growth of NiZn and MnZn-ferrites films at relatively low temperature (200 ºC). This is likely because Zn ions readily form oxygen tetrahedrons at a lower temperature. Similarly, in the case of $Fe_3O_4$, simultaneous stabilization of $Fe^{2+}$ and $Fe^{3+}$ oxidation states requires that the films be grown under fairly reducing conditions (~$10^{-6}$ Torr) and at low substrate temperatures (~350 ºC).[20] For NFO and CFO, which are not easily oxidized, films are generally grown and/or ex-situ annealed at relatively high temperatures ($\geq$ 600 °C).

To realize epitaxial growth of NFO and CFO films at lower temperatures, one needs to enhance the oxidation ability and also the kinetic energy of the ablated species. For conventional PLD in oxygen ambient, the background pressure used is generally in the range of several hundred mTorr. In this case, the mean free path of molecules is much shorter than the target-substrate distance and the kinetic energy of ablated species is dramatically reduced because of gas phase scattering and reaction.[21] The oxidation capability at lower temperatures can be significantly enhanced by using ozone, thereby also enabling reduction of the background pressure required for stoichiometric film growth. If the pressure can be reduced to a few mTorr while maintaining the oxygen activity, the mean free path of molecules will be comparable to the target-substrate distance and the high kinetic energy of the ablate species reaching the substrate can be preserved to enable film growth at lower temperatures. In this study, we report a robust epitaxial growth mode for spinel ferrite films which can overcome the obstacles incurred in conventional film growth mode. Specifically, we demonstrate that epitaxial Ni/Co spinel ferrites films, with large-scale atomically flat surface and enhanced magnetic moment, can to be grown at significantly lower temperatures than previously reported.

## II. Experimental

We prepared a series of NFO and CFO on (100)-oriented MAO substrates at various temperatures (175, 250, 325, 400, 550, 690 ºC), all under identical laser conditions of near 1.5 J/cm$^2$ and high repetition rate of 10 Hz, and with background oxygen pressure of 10 mTorr mixed with 10-15% of ozone. The film growth was monitored *in situ* using reflection high energy electron diffraction (RHEED). After growth, the samples were cooled down to room temperature under an ozone/oxygen pressure of 100 mTorr. Prior to deposition, the substrates were annealed between 1300-1400 ºC in air for 6 hrs to obtain an atomically-flat step and terrace profile. NFO films were also grown at temperatures as low as 100 ºC but they showed poor magnetic properties, although the surface morphology and RHEED pattern were similar to higher temperature grown films. Films grown at room temperature were amorphous. For comparison purpose, some films were also deposited on (100)-oriented STO and MgO substrates at 250 ºC.

A standard θ-2θ x-ray diffraction setup (Phillips X' pert Pro) was used to determine the phase and epitaxy of the films. XRD measurements were performed using a CuKα source operating at 45kV and 40 mA. Films surface morphology was characterized using atomic force microscopy (Veeco NanoScope) scanned in the tapping mode. The growth rate was calibrated using X-ray reflectivity, and film thicknesses were measured by Dektak surface profilometer and AFM. Energy-Dispersive X-ray Spectroscopy (EDS) was used to determine the film cation stoichiometry. Magnetic properties of the NFO and CFO films were measured using a superconducting quantum interference device (SQUID) magnetometer.

###    III.     Results and Discussion

Atomic force microscopy (AFM) measurements reveal that all films grown at and below 550 ºC on MAO substrate are atomically flat with clear steps, virtually identical to the substrate - even for films thicker than 200 nm, as shown in Fig. 1(a) and (b) for NFO films. Typical step heights are nearly 4 Å (half unit-cell) or 8 Å (one unit-cell), as shown in Fig. 1(d). On increasing the temperature to 690 ºC, quasi-3D growth is observed for the NFO film as shown in Fig. 1(c), but the step pattern of the substrate can still be faintly distinguished. A $1\times1$ μm$^2$ image of the film is provided in the inset to Fig. 1(c) exhibiting islands of less than 3 nm in height (Fig. 1(e)). Almost identical surface morphology has also been observed as a function of growth temperature for CFO films (not shown). The AFM measurements are consistent with the RHEED patterns recorded after films growth, as shown in the upper-left insets of Figures 1(a) and 1(c). All the films (except the film grown at 690 ºC) show very clear streaky RHEED patterns, indicating two-dimensional (2D) type epitaxial growth. No RHEED oscillations are observed, but the RHEED pattern and streak intensities remain essentially unchanged throughout the growth process, similar to that observed for step flow growth mode. Moreover, we have determined that the surface morphology, as determined using AFM, remains essentially the same for films with different thicknesses. Films grown at 690 ºC exhibit a spotty RHEED pattern, indicating island growth (3D). We have also grown NFO and CFO films on STO and MgO substrates at 250 ºC, all of which show atomically flat surfaces, essentially identical to the substrate morphology as shown in Fig. 2, suggesting that the growth mechanism is independent of the substrate. We have used Energy-Dispersive X-ray Spectroscopy (EDS) to determine the film stoichiometry. For all the NFO and CFO films, the Fe to Ni/Co ratio is found to be close to the expected value of 2.0 based on the target composition.

We have carried out detailed texture and epitaxy analysis of our films, including symmetric 2θ-θ scans, rocking curves, phi scans and reciprocal space maps. Large-angle 2θ-θ scans show only diffraction peaks corresponding to NFO or CFO film and the MAO substrate. No evidence of any secondary phases is found. In Fig. 3(a) and (b) we plot $\theta$ - $2\theta$ spectra near the (400) reflection of the MAO substrate of NFO and CFO films, respectively, grown at different temperatures. All the films exhibit clear x-ray diffraction peaks corresponding to the (400) film reflections, with the peak position gradually shifting to higher angles with increasing substrate temperature, indicating a decrease in the out-of-plane lattice parameter ($a_z$). This can either be due to increasing strain relaxation or reduced oxygen deficiency with increasing growth temperatures. Either way, since all films are nominally of the same thickness (within 5% of each other), this effect is completely temperature driven.

In Fig. 3(c) we plot the growth temperature dependence of the calculated $a_z$ values. As is clear from the plots, NFO and CFO films behave quite differently. For NFO films, $a_z$ decreases almost monotonically from 175 ºC to 690 ºC but still above the bulk value (8.340Å, blue line) even at the highest growth temperature ($a_z$ = 8.361Å at 690 ºC). For CFO films, $a_z$ increases much more rapidly for films grown below 325 ºC. At higher temperatures, the lattice constant appears to saturate to a value of nearly 8.405 Å, somewhat above the bulk value of 8.390 Å for CFO (red line).

Epitaxial analysis has been further carried out by performing phi-scans around the (311) reflection of the film and substrate. In Fig. 3(d) we show the phi-scans for the NFO films (CFO films show qualitatively similar behavior). Four sharp peaks, 90 degrees apart, confirm cubic symmetry of the films down to the lowest temperature (175 ºC). Higher temperature improves both the FWHM and peak intensities of the phi scans, consistent with the texture analysis from rocking curves measurements as shown in Fig. 4. We observe more than a factor of five improvement in full-width at half-maxima (FWHM) and intensity for NFO films grown at 550 ºC (FWHM ~ 0.3º) and 690 ºC (FWHM ~ 0.2º) as compared to the low temperature films (FWHM ~ 1.0º). For the CFO films, the improvement in texture is more gradual as compared to NFO.

Overall, x-ray measurements reveal varying degree of epitaxy for all films, with increasing temperature improving both the film texture and epitaxy. This conclusion, even though understandable is still surprising for the following reason: One straight-forward consequence of low temperature growth is the lack of surface diffusion for the incident species, and at a low enough temperature this is expected to prevent single phase formation and, therefore, crystalline growth. This clearly is not the case for our films. Since we do not find any evidence of temperature-related phase instability, we conclude that the temperature requirement for the NFO and CFO phase formation is lower than 175 ºC, particularly in the phase space of our deposition conditions. Once phase formation is guaranteed, epitaxy and texture improves steadily with temperature, as expected. Also, the epitaxy and texture improvement can be roughly correlated with smoothening of the step-terrace features of the AFM images at higher temperatures. We next present our magnetic analysis.

In Fig. 5(a) we plot magnetization loops measured at 5K for NFO (blue) and CFO (red) films grown at 325 ºC. Films grown at other temperatures exhibit a similar loop shape and coercivity, differing only in the saturation magnetization. The magnetization saturates fairly well at around 3 Tesla, suggestive of a low density of anti-phase boundaries (APBs), consistent with the report of Rigato et al.[22] The coercive fields are about 0.3 T and 1.1 T for NFO and CFO films, respectively. These values are higher than the literature reports, but it is well known that coercivity depends on extrinsic factors such as growth conditions, post-deposition annealing, etc. Fig. 5(b) shows the growth temperature dependence of the saturation magnetization ($M_S$) values of NFO and CFO films. For both films, a similar qualitative trend is observed. The $M_S$ values are close to the bulk values of 300 emu/cc and 450 emu/cc for NFO and CFO, respectively, for the films grown at higher temperatures (550 ºC and 690 ºC). Among the films grown at lower temperatures, we observe an enhanced magnetization for films grown at 400 ºC and 325 ºC, while the values are somewhat below that for the bulk for films deposited at or lower than 250 ºC. The increase in the saturation magnetization value of CFO films grown at intermediate temperatures is especially striking, over 500 emu/cm$^3$ - about 15-20% higher than the bulk value. As mentioned earlier, such anomalous values have previously been reported in ultra-thin NFO films.[5] In our films, the high magnetization values are also likely due to temperature-induced cation-disorder. It is encouraging that even for films grown at 175 ºC the $M_S$ values are still close to the bulk values implying even these films are not far from the expected inverse-spinel configuration. The NFO and CFO films grown on STO substrates show reduced saturation magnetization of nearly 200 emu/cc for NFO, and less than 300 emu/cc for CFO. This is consistent with earlier literature reports.[14]

Here we discuss the possible growth mechanism. As illustrated in Fig. 6(a), the (100) spinel surface has a relatively open structure with both the octahedral and tetrahedral surface sites not being blocked during the growth process. The tetrahedral sites form a ($\sqrt{2}$/2)$a_s$×($\sqrt{2}$/2)$a_s$-45º ($a_s$ = lattice constant) unit cell at the surface (marked as black square in Fig. 6(a)) and the octahedral sites (marked as red rectangle) forms a ($\sqrt{2}$/2)$a_s$×($\sqrt{2}$/4)$a_s$-45º unit cell. For the inverse spinel ferrites NFO and CFO, the Fe ions will occupy both the octahedral and tetrahedral sites, while Ni/Co ions will occupy only the octahedral sites. The relatively low ozone/oxygen pressure used during growth ensures that the incident species have sufficient kinetic energy required for such an activity. Furthermore, ozone helps in enhancing the oxidation kinetics at low temperatures. We speculate that the ablated cation species that reach the substrate surface locally crystallize at the nearest available site without requiring long-range surface diffusion. This, we believe, is unique to spinels and maybe related to the fact that many equivalent configurations are possible that maintain the spinel crystallographic symmetry. We term this growth as "vertical step-flow mode" since it is reminiscent of the "lateral" step-flow growth where the adatoms nucleate at the step terrace and propagate to the step edge (see Fig. 6 (b)). This helps explain why the streaky RHEED pattern remains essentially unchanged during growth and atomically flat surfaces are preserved even for thick films.

## IV. Summary

In summary, we have demonstrated a novel growth mode, which we term "vertical step-flow", for spinel ferrite films that results in high quality films with large-scale atomically flat surface and enhanced magnetic moment. The growth mode is independent of the nature of substrates (MAO, STO, MgO) and film thicknesses, and results in epitaxy at significantly lower temperatures than previously reported. While the underlying mechanism is not completely understood, we speculate that our study will inspire further investigation for this novel growth process and promote device applications of spinel films.

**Acknowledgements**

This work was supported by ONR (Grant No. N00014-09-0119). We thank G. Srinivasan of Oakland University for helpful discussions and suggestions.

**Figure Captions**

**Fig. 1.** 10 µm×10 µm AFM images of NFO films grown at (a) 250 ºC, (b) 550 ºC, and (c) 690 ºC on MAO (100) substrates. Upper-left insets in (a) and (c) are the RHEED patterns recorded at the completion of film growth. The lower-left inset in (a) is the typical morphology (10 µm×10 µm) of a high temperature annealed MAO (100) substrate. The lower-left inset in (c) is a zoom-in of the film surface (1 µm × 1 µm). Fig. 1(d) and (e) are line profiles as marked in (a) and (c).

**Fig. 2.** 2×2 µm$^2$ AFM image of a 120 nm NFO thin film grown at 250 °C on (001) STO substrate. Clear step and terrace features of height 4Å are observed, similar to the MAO substrate. Lower-left inset shows the substrate morphology.

**Fig. 3.** X-ray diffraction θ-2θ scans of (a) NFO and (b) CFO films grown at different temperatures. (c) Growth temperature dependences of the determined lattice constants in the film normal direction. Solid lines are the bulk NFO/CFO lattice constants. (d) Phi scans of NFO films grown at different temperature.

**Fig. 4.** Full width at half maxima (FWHM) values from the rocking curve measurements and phi scans of NFO films grown at various temperatures. All the samples are between 200-220 nm thick. The films grown at temperatures of 550 °C and 690 °C exhibit better epitaxy and texture as compared to the low temperature films. CFO films also show a similar (but slower) trend with increasing temperature.

**Fig. 5.** (a) Typical magnetic hysteresis loops of NFO and CFO films grown at 325 ºC, measured at 5K with magnetic field applied along the [001] in-plane direction. (b) Plot of the saturation magnetization of NFO and CFO films versus growth temperature.

**Fig. 6**. (a) Surface structure of AB$_2$O$_4$ spinel (001) with three layers illustrated. The black square represents the tetrahedral lattice sites and the red rectangle represents the octahedral sites. (b) and

(c) illustrate the propose "vertical" and its comparison with the well established "lateral" step-flow growth mode, respectively.

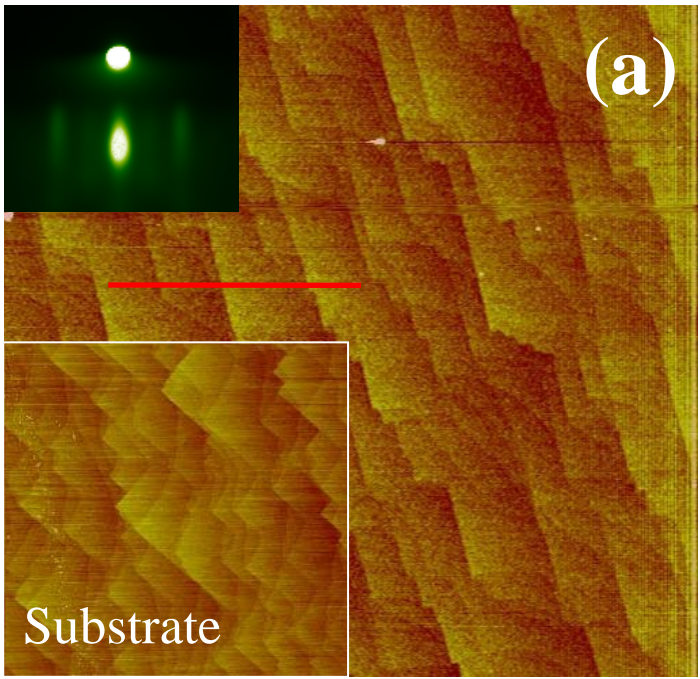
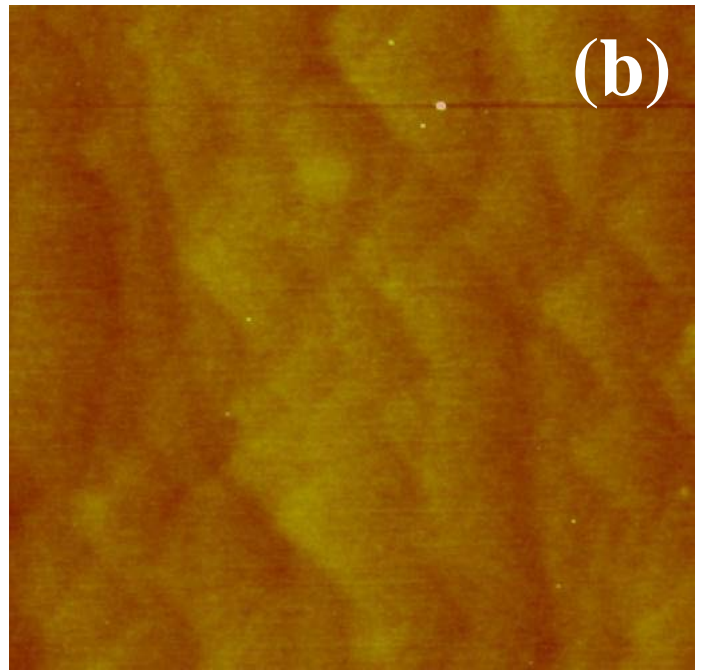
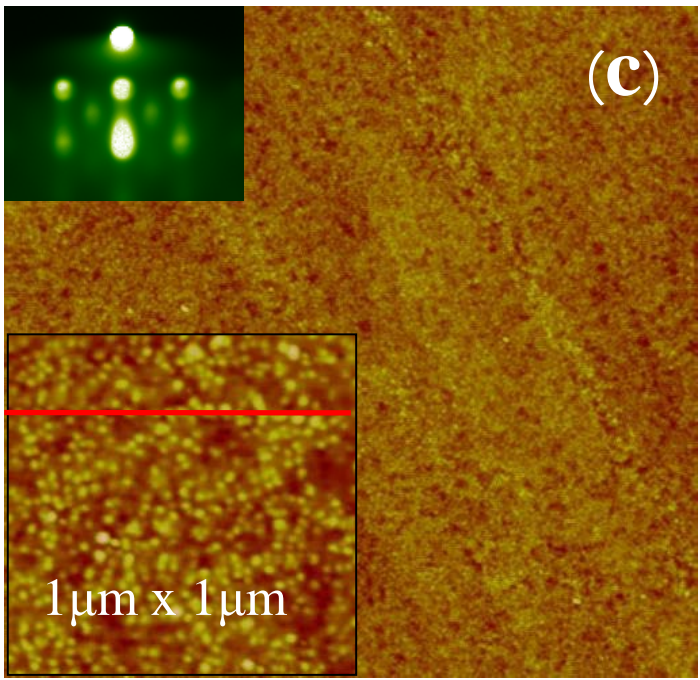
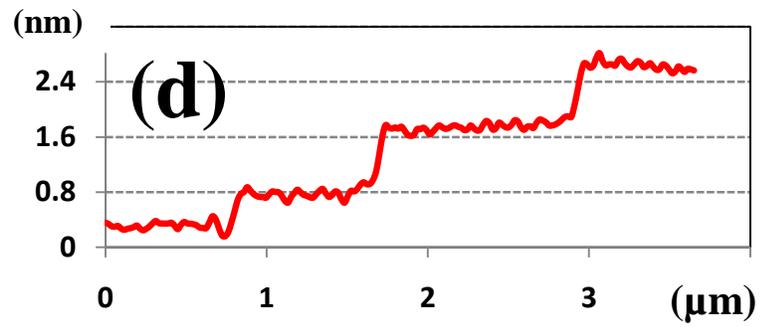
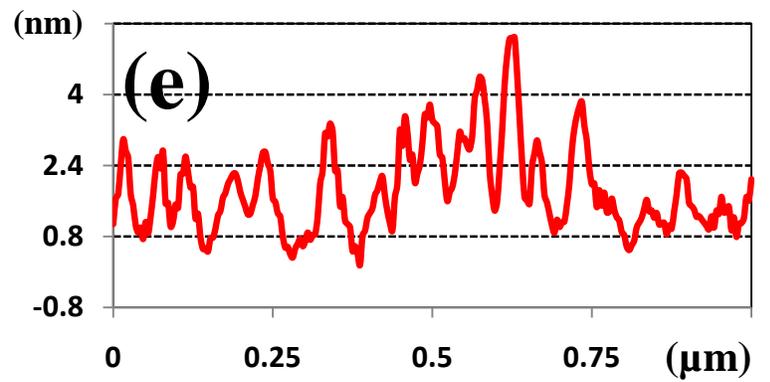

Ma et al, Fig. 1

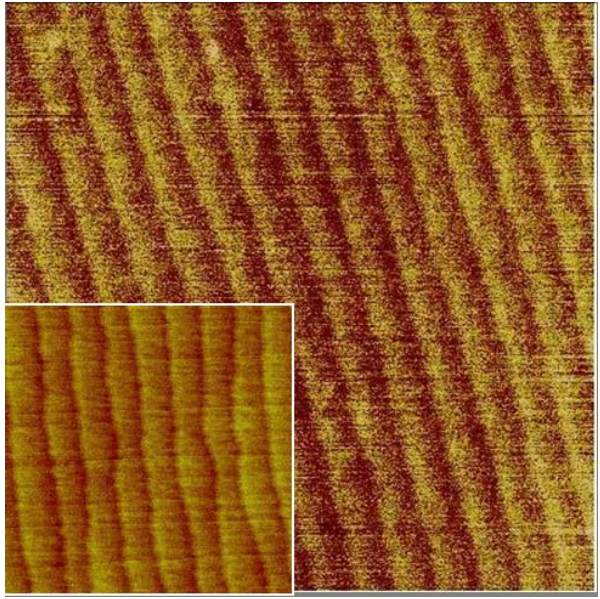

Ma et al, Fig. 2

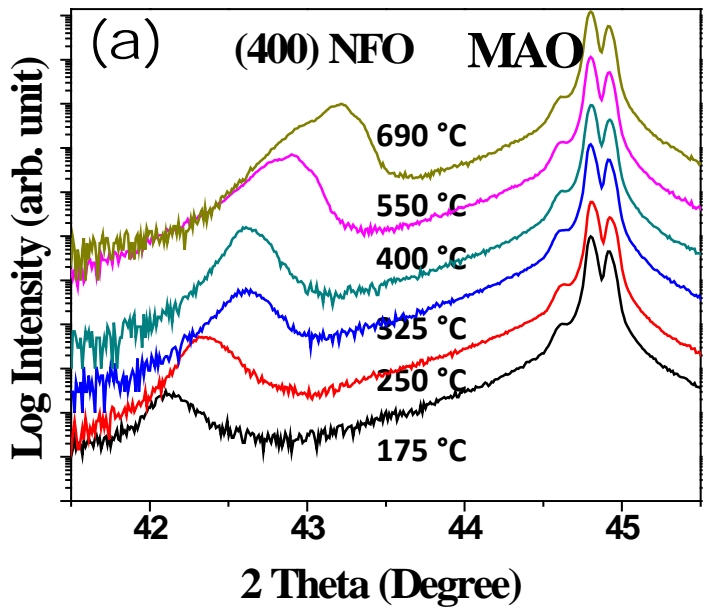
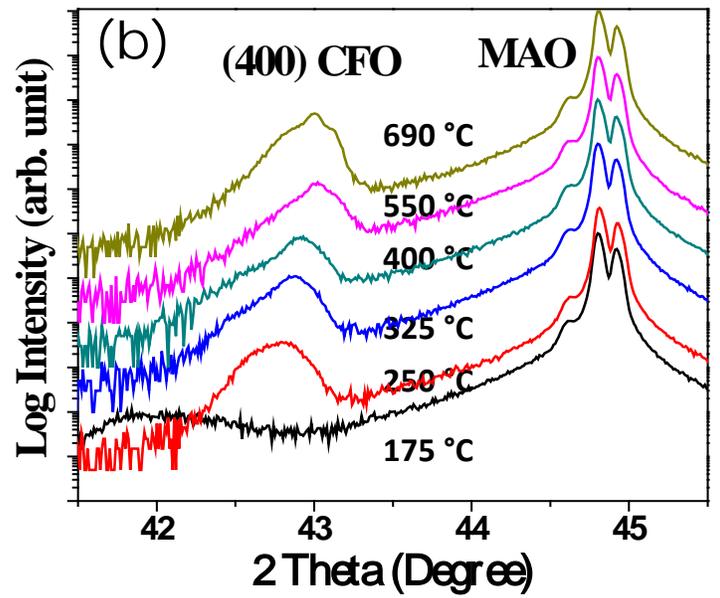
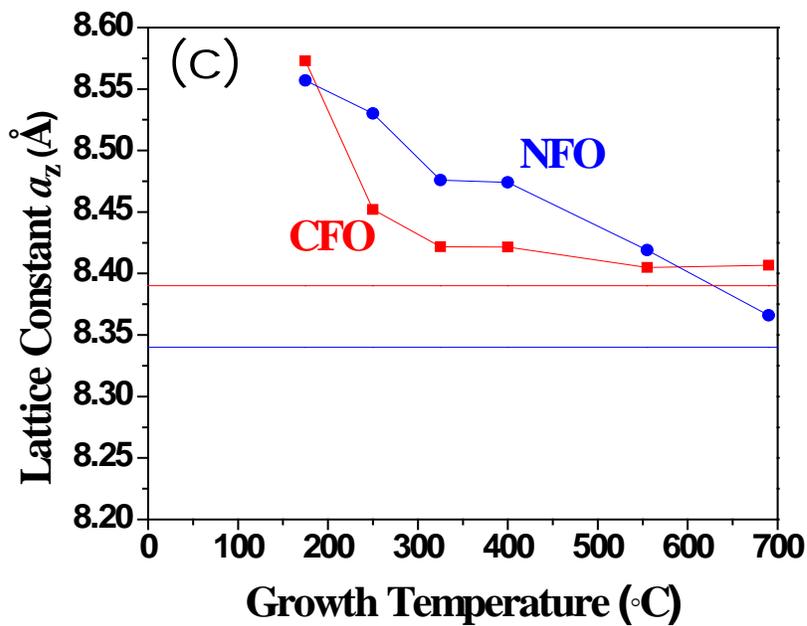
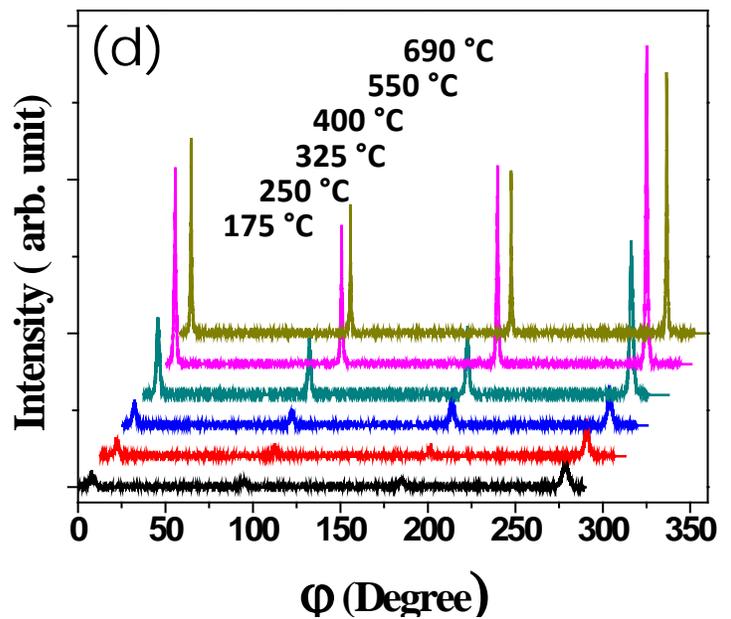

Ma et al, Fig. 3

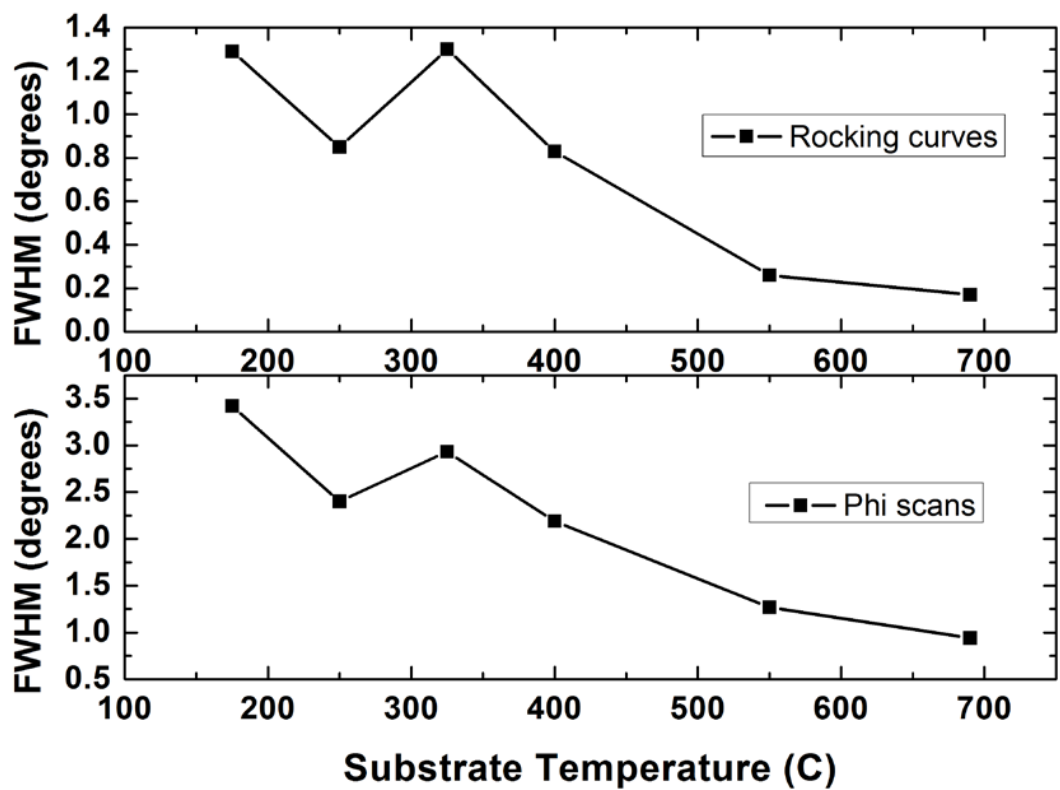

Ma et al, Fig. 4

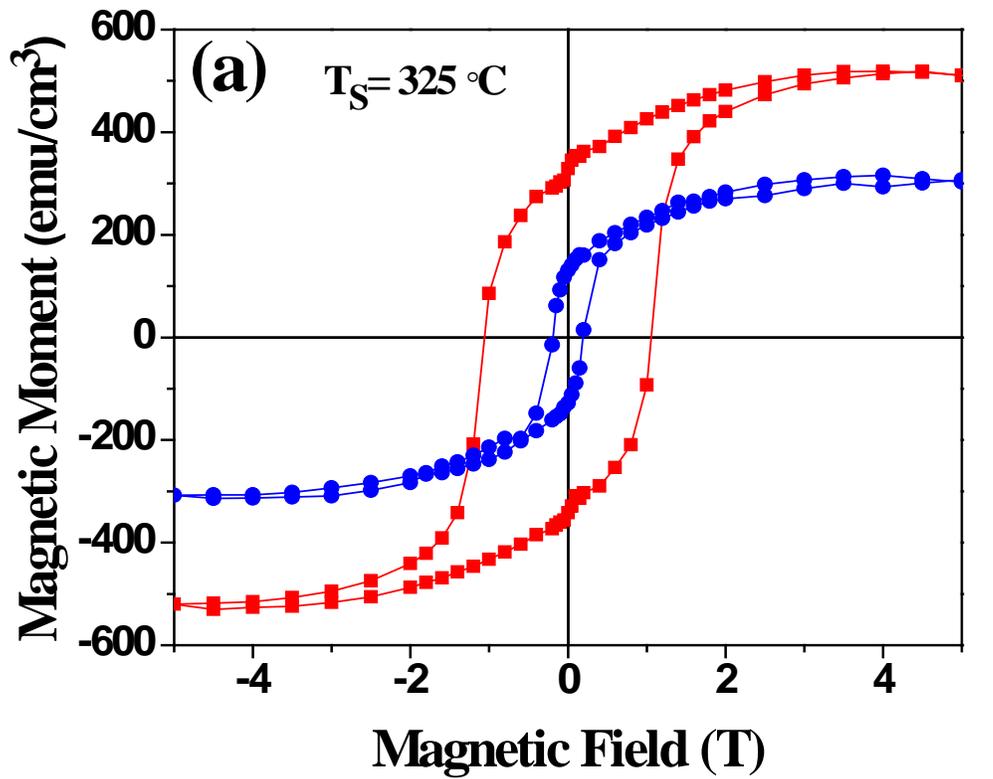

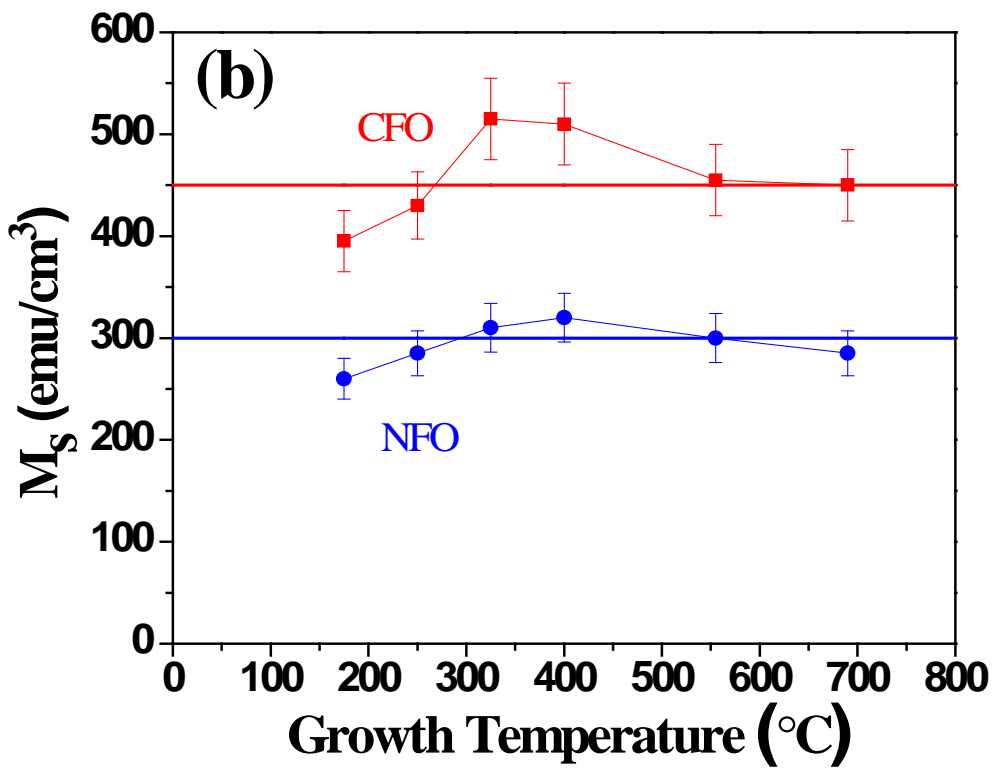

Ma et al, Fig. 5

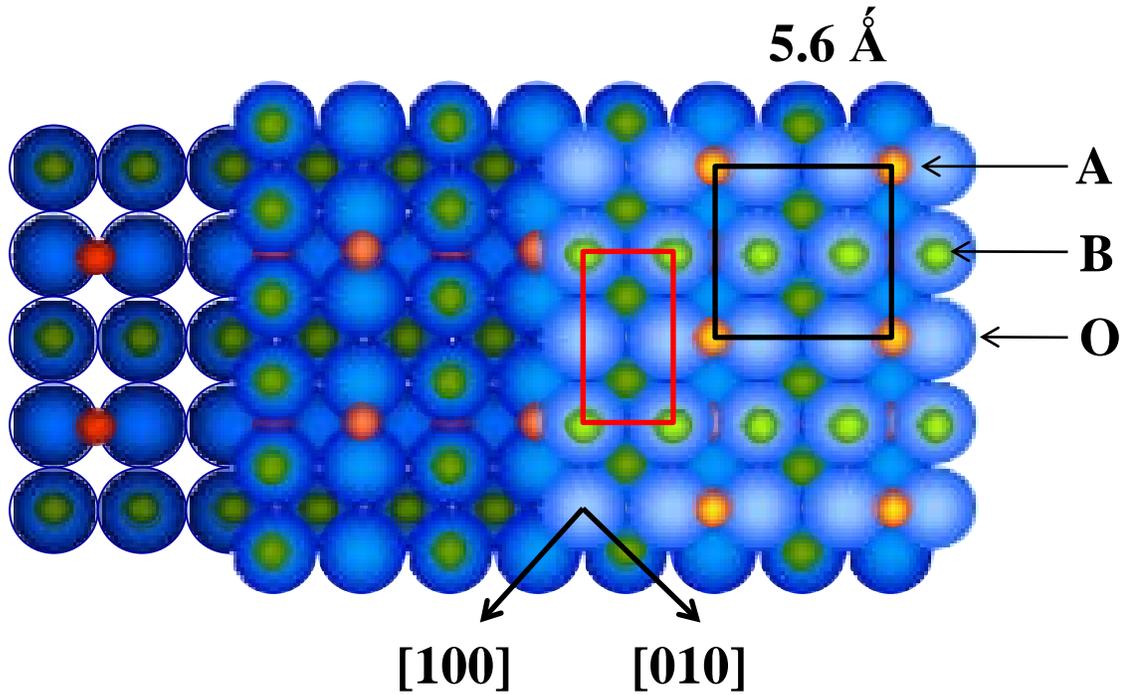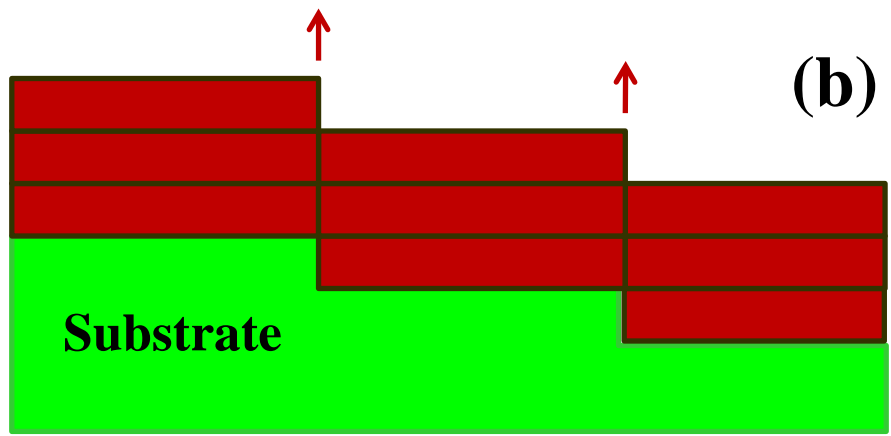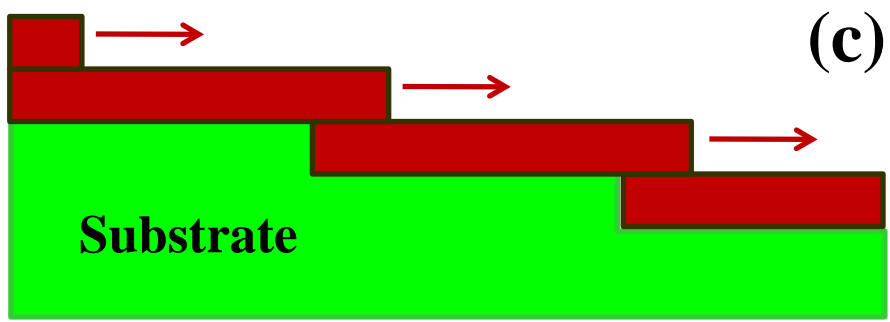

Ma et al, Fig. 6